\pdfoutput=1

\documentclass[11pt]{article}

\usepackage[final]{acl}

\usepackage{times}
\usepackage{latexsym}
\usepackage{booktabs}
\usepackage{multirow}
\usepackage{makecell}
\usepackage{svg}
\usepackage{graphicx}
\usepackage{subcaption}
\usepackage{amsmath}
\usepackage{algorithm}
\usepackage{algpseudocode}
\usepackage{ulem}
\usepackage{amsfonts} 

\usepackage[T1]{fontenc}

\usepackage[utf8]{inputenc}

\usepackage{microtype}

\usepackage{inconsolata}

\usepackage{graphicx}

\newcommand{\unigram}{\texttt{UNIGRAM-WATERMARK} }
\newcommand{\sir}{\texttt{SIR} }
\newcommand{\sirr}{\texttt{SIR}}

\newcommand{\measurement}[2]{%
    #1{\scriptsize$\pm#2$}%
}

\title{Revisiting the Robustness of Watermarking to Paraphrasing Attacks}

\author{Saksham Rastogi \\
  Indian Institute of Science \\
  Bengaluru, India \\
  \small{\texttt{iitdsaksham@gmail.com}} \\\And
  Danish Pruthi \\
 Indian Institute of Science \\
  Bengaluru, India \\
  \small{\texttt{danishp@iisc.ac.in}} \\}

\begin{document}
\maketitle

\begin{abstract}

Amidst rising concerns 
about the internet being proliferated 
with content generated from language models (LMs), 
watermarking is seen as a principled way
to certify whether text was generated from a
model.
Many recent watermarking techniques
slightly modify  
the output probabilities 
of LMs to embed a signal in the generated output 
that can later be detected.
Since 
early proposals for text watermarking, 
questions about their robustness to paraphrasing 
have been prominently discussed. 
Lately, 
some techniques are 
deliberately designed 
and claimed to be robust to paraphrasing.
However, such watermarking schemes 
do not adequately 
account for the ease 
with which they 
can be reverse-engineered.
We show that with access to only a limited number of generations from a black-box watermarked model, we can drastically increase the effectiveness of paraphrasing attacks to evade watermark detection, thereby rendering the watermark ineffective.\footnote{Data and code to replicate our evaluation is available at: 
{\fontfamily{qcr}\selectfont \href{https://github.com/codeboy5/revisiting-watermark-robustness}{https://github.com/codeboy5/revisiting-watermark-robustness}}}

\end{abstract}

\section{Introduction}

Given the remarkable 
fluency and relevance with 
which language models (LMs) respond
to varied queries, it is  
challenging for humans to distinguish 
language model outputs from human-written text. 
Past studies note that 
human performance in making such a distinction 
is close to that of random chance~\cite{gehrmann2019gltr, brown2020language}. 
In response, watermarking language models 
is seen to be a principled way to certify 
whether a piece of text was generated by a model.

A prominent watermarking 
approach
works by 
implanting a signal during decoding, 
wherein a certain set of tokens (aka a green list) is boosted~\cite{kirchenbauer2024watermark}. 
This signal, albeit imperceptible to an unsuspecting 
reader, can be verified by running a statistical test. 
For watermarking to be effective, 
the implanted signal should be easy to detect and hard 
to remove.
Unsurprisingly, 
there has been considerable 
discussion about the robustness of 
watermarking approaches 
against paraphrasing attacks~\cite{krishna2023paraphrasing, kirchenbauer2023reliability}. 

There exist different ways of choosing tokens in the green list and the extent to which they should be boosted.
These approaches offer varying levels of robustness against paraphrasing.  
The original paper~\cite{kirchenbauer2024watermark} recommends 
pseudo-randomly selecting a different set of green tokens
at every timestep  
based on a hash of the last $k$ tokens. The 
authors note that 
higher values of $k$ 
would render the watermarking scheme ineffective, as any changes to a token would disrupt the green lists for the next $k$ timesteps, and therefore suggest using the last one or two tokens ($k$~=~$1$ or $2$).\\
A recent study~\cite{zhao2023provable} argues that ``a consistent green list 
is the most robust choice,'' as any modifications to the input text have no effect whatsoever on the (fixed) green list.
Relatedly, ~\citet{liu2024semantic} propose 
a ``semantic-invariant robust'' watermarking  (\texttt{SIR})
which is 
designed to 
produce similar green lists 
for semantically-similar contexts 
and is touted to be robust to paraphrasing. %

In 
this ongoing debate, our work highlights 
just how easy it is to decipher the green list 
for both the semantic-invariant watermarking scheme~\cite{liu2024semantic} and watermarking with consistent green list~\cite{zhao2023provable}.
While a recent contemporaneous study~\cite{jovanovic2024watermark} 
corroborates 
that watermarking with a fixed green list can 
be easily reverse-engineered,  
we show that similar results also hold for semantic-invariant watermarking scheme from \citet{liu2024semantic}. 
For both these watermarking schemes,
with just $200$K tokens of watermarked output, 
we can predict green lists with over $0.8$ F1 score. 
This knowledge of green lists can be exploited 
while paraphrasing to launch attacks that cause the detection rates to plummet below $10$\%, rendering the watermark ineffective. 

Overall, our findings suggest that one should consider the possibility of reverse-engineering the watermarking scheme, 
when discussing its robustness to paraphrasing attacks. 
Our work also 
raises potential concerns about the generalization of watermarking algorithms that use machine learning models to generate the watermarking signal. 

\section{Background}

A prominent approach to watermarking is to compute watermarking logits that are added to logits generated by a language model at each generation step.
Formally, for a language model $\mathcal{M}$ 
with vocabulary $V$, 
and a prefix comprising tokens $\mathbf{w}_1, \mathbf{w}_2, \ldots, \mathbf{w}_n$, 
the scheme involves first computing the logits
 $\mathcal{M}(\mathbf{w}_1 \ldots, \mathbf{w}_n) = (l_1, \ldots, l_{|V|})$ 
 of the language model that would ordinarily be used to predict the 
subsequent token. 
As per \cite{kirchenbauer2024watermark}, 
the last $k$ tokens,   
$\mathbf{w}_{n-k+1}$ to $\mathbf{w}_{n}$,
are then fed to a psuedo-random function $F$
to partition $V$ into a green list $G$
 and a red list  $R$
such that $|G| + |R| = |V|$.
Finally, the logits 
corresponding to the tokens in the green list, $G$, are boosted
by $\delta$ ($\delta > 0$). %
The watermark can then be detected through a one-proportion z-test on the fraction of green tokens in the generated text.

A recent study \cite{zhao2023provable}
makes a case for using a fixed green list 
(where the partitioning function, 
$F$, 
does not depend on the context)
to confer robustness to paraphrasing attacks.
The underlying intuition 
is that any changes in 
the text will not disrupt the 
constant green list. 
To counter paraphrasing attacks, 
another recent proposal~\cite{liu2024semantic}
is to train a model, $\mathcal{W}$, 
to output watermarking logits
using the context:
$\mathcal{W}(\mathbf{w}_1 \ldots, \mathbf{w}_n) = (\delta_1, \ldots, \delta_{|V|})$.
This model, $\mathcal{W}$, is designed
such that similar contexts yield 
similar watermarking logits. 
This property is supposed to 
make models robust to paraphrasing. Further, diverse contexts 
are supposed to yield 
different watermarking logits, thus making it hard to reverse-engineer the green list---this is not true in practice, as we show later in our experiments (\S\ref{exp:results}). %

\paragraph{Paraphrasing Attacks.} 
\citet{krishna2023paraphrasing} introduce a controllable paraphraser and launch paraphrasing attacks on various text detectors. Their findings indicate that although paraphrasing reduces the effectiveness of most AI-generated text detectors, watermarking is the most resilient method. Another study \cite{kirchenbauer2023reliability} investigates the reliability of watermarks across different paraphrasing models and suggests that the reliability of watermarking should be discussed in terms of the length of the available input. The study concludes that watermarking is extremely reliable for longer texts.

\section{Methods} \label{methods}

We study the robustness 
of watermarking approaches
against paraphrasing attacks.
Unlike prior attacks~\cite{krishna2023paraphrasing, kirchenbauer2023reliability}, 
we first
attempt to decipher 
the tokens in the green list 
and then
incorporate 
that knowledge in existing paraphrasing attacks.

\subsection{Estimating Green Lists}

We assume access to only generations from the watermarked model, with no access to model weights or its tokenizer.
To decipher the green list,  we use a simple counting-based algorithm similar to the ones used in prior work \cite{zhao2023provable, sadasivan2024aigenerated}. Specifically, we compare the relative frequencies of tokens in a corpus generated by the watermarked model against their relative frequencies in a corpus of human-written text. 
Tokens that exhibit a higher relative frequency in the watermarked corpus compared to the reference corpus are classified as green tokens.
We present the detailed algorithm in Appendix \ref{sec:appendix}. 

This approach is similar to the one proposed in a contemporaneous work \cite{jovanovic2024watermark}, where for each token, they compute two conditional probabilities: probability of a token given its preceding context in a watermarked corpus and the same probability in a base 
corpus. They investigate two scenarios for obtaining the base corpus: using a non-watermarked version of the same language model or using a different language model as a proxy for the base distribution. 
In contrast, our approach does not require access to the unwatermarked language model for the base distribution; instead, we derive our base distribution from the OpenWebText corpus.\footnote{
To demonstrate the robustness of our algorithm across different base distributions, we present additional results in the appendix, where we estimate the green list using the RealNewsLike subset of the C4 dataset \cite{JMLR:v21:20-074}.} Furthermore, our approach assigns binary scores of 0 and 1 for tokens in the red and green lists, respectively. 
Please note that the green list once estimated using the algorithm can be used to launch paraphrasing attacks on a variety of downstream datasets (details in \S\ref{exp:dataset}).

\paragraph{A note about metrics:} 
Prior work relies on F1 score to evaluate the correctness of predicting the green list. 
However, this metric assumes equal importance for all tokens and fails to account 
for the fact that natural language follows a Zipf's law, wherein the frequency of a word is inversely proportional to its rank in the list (sorted in decreasing order of word frequencies). 
While it may seem like a minor technicality,  
we show that the traditional F1 score  
overestimates the security of watermarking.

To address this limitation, we suggest using a generation-based F1 score that 
 computes the F1 score for classifying tokens into green or red list for each token \emph{in text generated from watermarked models}. This small change incorporates the relative frequency of each token.

\subsection{Paraphrasing with Green Lists}
\label{methods:green_list}

One can imagine that incorporating prior knowledge about green lists 
should be able to improve 
the efficiency of off-the-shelf paraphrasers to remove the watermark signal and evade detection.  
Since many paraphrasing models 
are also auto-regressive
generative models~\cite{krishna2023paraphrasing, lewis2020pretraining, lin2021documentlevel, Witteveen_2019}, 

one can introduce an inverse watermarking signal into the generated text. Specifically, at every generation timestep, we subtract a small positive $\delta$ from the logits corresponding to tokens predicted to be in the green list.

\section{Results \& Discussion}

We first share details about our setup and then discuss the results of paraphrasing attacks.

\subsection{Experimental Setup} 
\label{exp:dataset}

We primarily consider two watermarking schemes that are designed and 
thought to be robust against paraphrasing, namely, the semantic-invariant robust (\texttt{SIR}) watermarking~\cite{liu2024semantic}  and watermarking with a fixed green list, which is referred as \texttt{UNIGRAM WATERMARKING} in a recent study that analyzes the robustness of this approach~\cite{zhao2023provable}.
We use the LLaMA-7B model~\cite{touvron2023llama} and apply the two watermark algorithms with hyperparameters of $\gamma = 0.5$ (fraction of green tokens) and $\delta=2.0$ (value used to boost the logits for green tokens) for all results presented in the main paper. Additional results using the Pythia model and other watermark hyperparameter choices are presented in 
Appendix \ref{sec:appendix_additional_results}. 

To evaluate the watermarking schemes and their robustness to paraphrasing, we use $50$-token prompts from Wikipedia articles \cite{wikidump} to generate $200$-token completions. (Note that this dataset is different from the one used to estimate the green list.)
We consider the subsequent $200$ tokens from the Wikipedia articles as human-written text for comparison. Additionally, we present results on prompts from arXiv papers \cite{cohan2018discourseawareattentionmodelabstractive} and Booksum \cite{kryściński2022booksumcollectiondatasetslongform} dataset in the Appendix (\S\ref{sec:appendix_additional_datasets}) to demonstrate the effectiveness of our attack on generations using prompts from diverse domains.

The results for each attack setting are aggregated across $500$ generations from the LLM. We use the DIPPER paraphrasing model  \cite{krishna2023paraphrasing} and incorporate 
the knowledge of (estimated) green list tokens (as described in \S\ref{methods:green_list}). 
To evaluate the detection accuracy of watermarking algorithms, we follow prior work and measure the True Positive Rate (TPR) at a low  False Positive Rate (FPR) of 1\% and 10\%. The False Positive Rate is set to low values to avoid falsely accusing someone of plagiarism.
We use the P-SP metric \cite{wieting2023paraphrastic} to assess the semantic similarity of paraphrases, past work considers the semantics of the paraphrase to be preserved if the P-SP value exceeds 0.76 \cite{krishna2023paraphrasing}. Additionally, we assess the quality of produced text by calculating the perplexity (PPL) using LLaMA-13B, which we consider as an oracle model.

\begin{figure}[t]
  \centering
  \includeinkscape[width=\columnwidth, height=8cm]{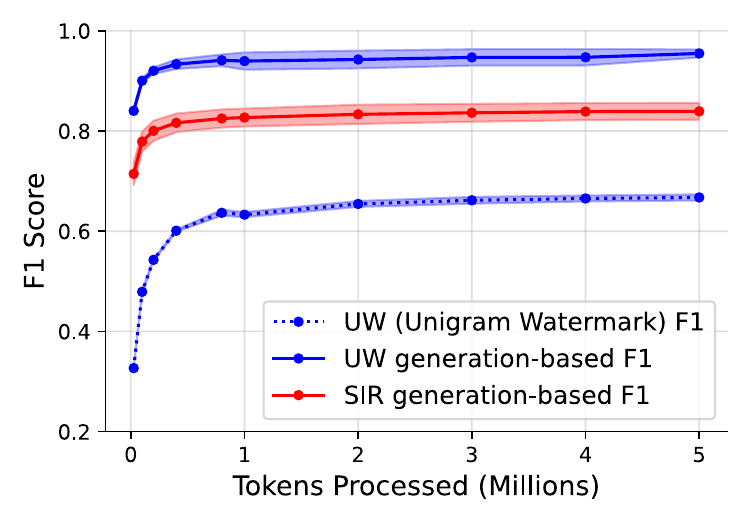}
  \caption{We show that with a limited amount of generated tokens, we can achieve a high F1 score for predicting the green lists of two watermarking schemes. 
}
  \label{fig:example}
\end{figure}

\begin{figure*}[ht]
    \centering
    \begin{subfigure}[b]{0.48\textwidth}
        \centering
        % \includesvg[scale=0.42]{figures/embVSwtm-iid.svg}
        \includeinkscape[scale=0.42]{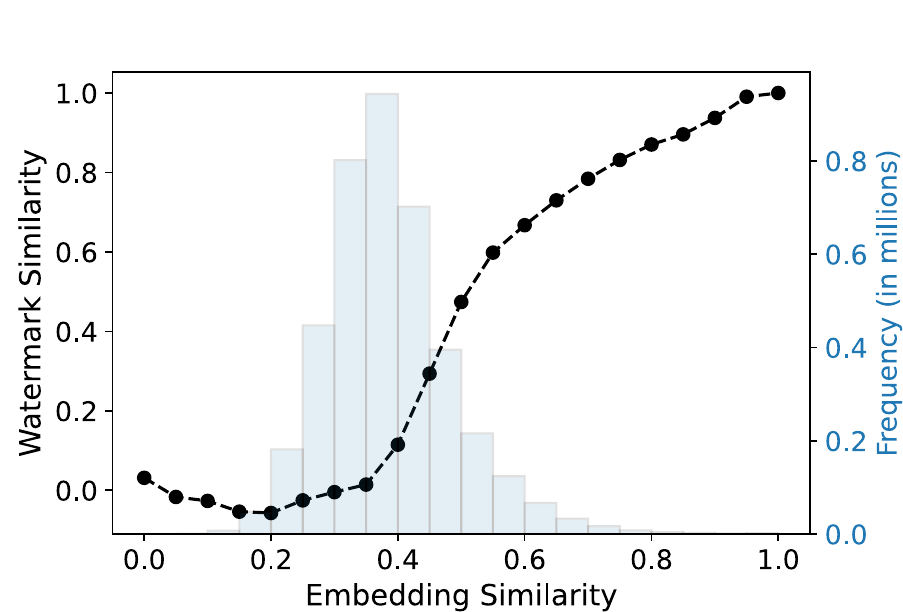}
        \caption{in-domain samples}
        \label{fig:plot1}
    \end{subfigure}
    \begin{subfigure}[b]{0.48\textwidth}
        \centering
        % \includesvg[scale=0.42]{figures/embVSwtm-ood.svg}
        \includeinkscape[scale=0.42]{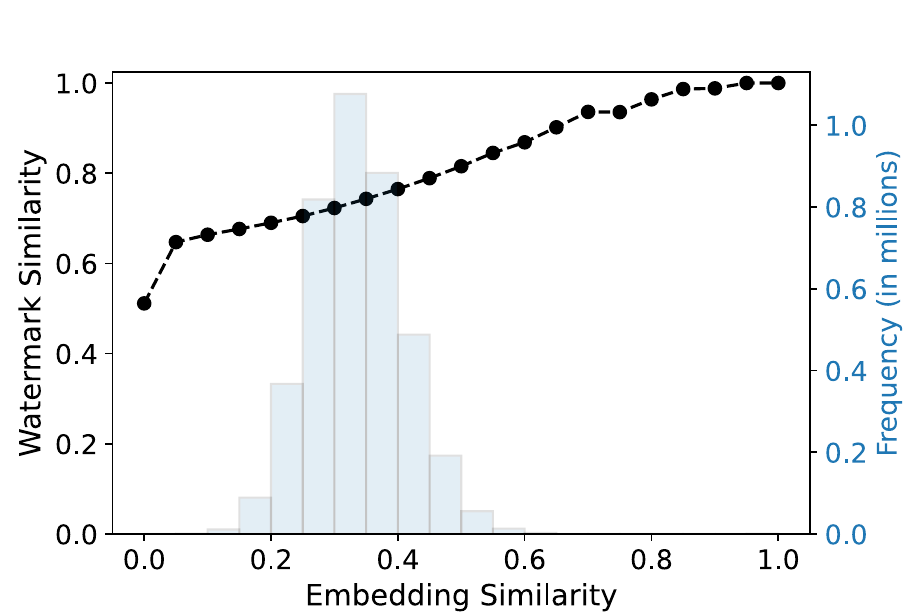}
        \caption{out-of-domain samples}
        \label{fig:plot2}
    \end{subfigure}
    \caption{For \sir watermarking, we depict
    the cosine similarity of the context embeddings (x-axis) vs the cosine similarity of the watermarking logits (y-axis). For in-domain samples, similar contexts produce similar watermarking logits and dissimilar ones produce different logits, however, this is not the case for out-of-domain samples. %
    }
    \label{fig:both_plots}
\end{figure*}

\subsection{Results} 
\label{exp:results}

\begin{table*}[h]
\centering
  \begin{tabular}{lcccccccc}
        \toprule
        & \multicolumn{4}{c}{\unigram} & \multicolumn{4}{c}{\sir} \\
        \cmidrule(lr){2-5} \cmidrule(lr){6-9}
        Attack & \makecell{TPR @\\ 1\% FPR} & \makecell{TPR @\\10\% FPR} & P-SP & PPL & \makecell{TPR @\\ 1\% FPR} & \makecell{TPR @\\ 10\% FPR} & P-SP & PPL \\
        \midrule
        No Attack & \measurement{99.3}{0.7} & 100.0 & 1.00 & \measurement{14.5}{1.0} & \measurement{93.3}{0.0} & \measurement{98.8}{0.1} & 1.00 & \measurement{12.8}{0.8} \\
        DIPPER (L20) & \measurement{88.7}{2.4} & \measurement{98.0}{0.5} & 0.95 & \measurement{11.3}{0.7} & \measurement{45.5}{0.9} & \measurement{82.7}{4.1} & 0.94 & \measurement{10.1}{0.6} \\
        DIPPER (L60) & \measurement{62.8}{2.2} & \measurement{92.1}{0.7} & 0.90 & \measurement{10.5}{0.7} & \measurement{24.0}{0.5} & \measurement{62.3}{2.4} & 0.90 & \measurement{9.6}{0.4}\\
        \midrule
        \textbf{Ours (L20)} & \measurement{3.2}{0.8} $\downarrow$ & \measurement{13.4}{2.5} $\downarrow$ & 0.87 & \measurement{11.6}{1.0} & \measurement{7.3}{2.1} $\downarrow$ & \measurement{20.3}{5.5} $\downarrow$ & 0.88 & \measurement{10.6}{0.5} \\
        \textbf{Ours (L60)} & \measurement{0.2}{0.2} $\downarrow$ & \measurement{1.9}{0.7} $\downarrow$ & 0.78 & \measurement{11.1}{1.1} & \measurement{3.8}{0.8} $\downarrow$ & \measurement{10.2}{3.1} $\downarrow$ & 0.81 & \measurement{10.2}{0.8} \\
        \bottomrule
    \end{tabular}
  \caption{We compare the detection rates of \unigram and \sir against paraphrasing attacks. We use two settings of the paraphrasing model, DIPPER, with lexical diversities (LD) of 20 and 60; higher LD implies stronger attack. Our attack involves modifying DIPPER with the estimated knowledge of the green list (details in \S\ref{methods:green_list}). We report the median P-SP \& PPL values.} 
  \label{table:paraphrase}
\end{table*}

We show that with just as few as $200$K tokens, \textbf{we can accurately predict whether a token belongs to green list} (Figure~\ref{fig:example}). 
It may be unsurprising that one can decipher the fixed green list used in the \texttt{UNIGRAM WATERMARKING}, as also documented by 
\citet{jovanovic2024watermark}. However, 
is noteworthy and 
surprising that even 
semantic-invariant watermarking (\texttt{SIR}) 
scheme, which dynamically produces a green list (based on the embeddings of the context)  
is just as vulnerable.

While the \sir approach aspires to 
produce similar watermarking logits for similar contexts and dissimilar ones for dissimilar contexts, we discover that this is not the case in practice. In Figure~\ref{fig:both_plots}, we plot the cosine similarity of the context embeddings vs the cosine similarity of the watermarking logits.
Interestingly, we notice that the aspired 
notion of 
producing similar watermarking logits for only similar contexts holds true only for in-domain samples and breaks for out-of-domain (OOD) samples. For OOD samples, the produced watermarking logits are highly similar regardless of the similarity in contexts (Figure \ref{fig:plot2}), suggesting that the green lists in \sir are not as dynamic as previously believed and are susceptible to be deciphered. Our findings suggest that other (future) watermarking algorithms that use machine learning to generate the watermarking logits might suffer from similar generalization concerns.

From Figure~\ref{fig:example}, we can also observe that the vanilla F1 scores present an overly optimistic picture about the security of watermarking approaches.
As discussed in \S\ref{methods:green_list}, the vanilla F1 metric weighs in all tokens uniformly. This approach fails to account for the long tail of rare tokens---whose presence in green or red list is hard to predict---which (by definition) occur infrequently in practical application. 
However, tokens that are actually generated can be predicted far more accurately, as can be clearly seen through about a $50$\% higher generation-based F1 score in Figure~\ref{fig:example}. %

Finally, we present results showing how the two watermarking schemes hold up against paraphrasing attacks (Table \ref{table:paraphrase}). We notice that the default DIPPER attack reduces the performance of both watermarking schemes. For FPR of $1$\%, it brings down the TPR to $88$\% (from $99.3$\%) for \unigram and to $45.5$\% (from $93.3$\%) for \sirr. 
When we 
empower the attack with the knowledge of (estimated) green lists,
the TPR values plummet to below $10$\%, rendering the watermarking schemes unusable. 
Across all setups, we confirm that the quality of LMs (measured through PPL) and the semantic meaning of paraphrases (evaluated via P-SP scores) is largely preserved.
Interestingly, our attack is slightly less effective for \sir than \unigram as our estimates for green lists are less accurate for \sirr.

\paragraph{A Note about Adaptive Text Watermark.} 
Just recently, an approach called Adaptive Text Watermark (ATW) was proposed,   
aiming to generate high-quality text while maintaining robustness, security, and detectability \citep{liu2024adaptivetextwatermarklarge}. Conceptually similar to \sirr, Adaptive Text Watermark generates a logit scaling vector ($v$) based on the semantic embeddings of previously generated text. 
The watermark is added to the LLM logits by proportionally scaling the original logits by a factor of ($1$ + $\delta . v$), where $\delta > 0$ controls the watermark strength. 
We find that 
while our attack strategy 
can significantly reduce the detection rate of ATW by about $10$\% (Table \ref{table:awd_attack}),
it is considerably 
more robust than the other two approaches.
Further analysis reveals that 
unlike SIR, the semantic mapping module of ATW (that converts embeddings of prefixes to logits) 
generalizes better for out-of-domain distribituions. 
This result suggests that semantics-based watermarking may be a viable alternate 
to defend against paraphrasing attacks, however, we suggest practitioners to confirm whether such learning-based approaches generalize to OOD domains.

\section{Conclusion}

We analyze watermarking schemes believed to be specifically robust to paraphrasing and show that it is easy to reverse engineer these algorithms and launch severe paraphrasing attacks. 
The effectiveness 
of our attacks 
underscores the need
to account for the ease of
reverse-engineering 
watermarking schemes, 
when discussing its robustness 
to paraphrasing attacks.
Additionally, 
we highlight that existing metrics concerning the security of watermarking are overly optimistic. 

\section{Limitations} 

Our work focuses specifically on watermarking schemes proposed to be robust against paraphrasing attacks. Future work can focus on other schemes, such as \cite{kuditipudi2024robust, aaronson2023reform}, which implant the watermark signal during sampling and claim to preserve the original distributions up to certain generation budgets. 
Another limitation of our work is that we do not address how effectively (ill-intentioned) humans can remove the watermark signal once they are aware of the estimated green lists. Additionally, paraphrasing attacks require significant compute as it uses a large language model for generating paraphrases.

\section*{Acknowledgements}

We thank the anonymous reviewers for their contructive feedback. We are also grateful to Anirudh Ajith for his feedback. 
DP acknowledges Adobe Inc., Google Research, Schmidt Sciences, National Payments Corporations of India (NPCI) and Pratiksha Trust for supporting his group’s research.

\bibliography{updated_refs}

\newpage
\appendix

\clearpage
\newpage
\section{Additional Experiment Details}
\label{sec:appendix}

\begin{table*}[ht]
\centering
  \begin{tabular}{lcccccccc}
        \toprule
        & \multicolumn{4}{c}{LLaMA 7B} & \multicolumn{4}{c}{Mistral 7B} \\
        \cmidrule(lr){2-5} \cmidrule(lr){6-9}
         & Precision & Recall & FPR & F1 & Precision & Recall & FPR & F1 \\
        \midrule
        \makecell{UW} & 0.89 & 0.46 & 0.05 & 0.61 & 0.91 & 0.51 & 0.05 & 0.65 \\
        \midrule
        \makecell{UW\\generation-based} & 0.96 & 0.88 & 0.08 & 0.92 & 0.98 & 0.92 & 0.05 & 0.95  \\
        \midrule
        \makecell{SIR\\generation-based} & 0.82 & 0.77 & 0.27 & 0.79 & 0.88 & 0.72 & 0.22 & 0.79  \\
        \bottomrule
    \end{tabular}
  \caption{We evaluate the performance of our watermark reverse engineering approach using a corpus of 1 million tokens, with the RealNewsLike subset of the C4 dataset serving as the base distribution. Our assessment metrics include precision, recall, false positive rate (FPR), and F1 score. These results indicate that our proposed paraphrasing attack is robust to the choice of the base distribution we use to reverse engineer the watermark scheme.}
  \label{table:c4_greenlist_results}
\end{table*}

\subsection{Algorithm for estimating the green list}

To estimate the green list, we compare the distribution of tokens between watermarked text and text from OpenWebText \cite{Gokaslan2019OpenWeb} dataset (to simulate the distribution of non-watermarked text). We query the LLaMA-7B watermark model with 50 token prompts from \cite{Gokaslan2019OpenWeb} to generate 256 token completions. We calculate the relative token frequencies for the watermarked text and text from the OpenWebText dataset. We use a minor modification of the algorithm used in \cite{zhao2023provable}, with the difference being using relative frequencies instead of absolute and using a small positive threshold $\tau$. We use a constant $\tau$ of $1 \times 10^{-6}$ across all our experiments.

$D_{\text{wtm}}$ and $D_{\text{human}}$ refer to the distribution of tokens in watermarked and non-watermarked text. %

\begin{algorithm}[h]
\caption{Estimating the Green List tokens}
\label{alg:estimate_green_list}
\begin{algorithmic}[1]
\For{every token $v$ in the vocabulary $\mathcal{V}$}
    \State $\Delta(v) \gets$ $D_{\text{wtm}}$($v$) $-$ $D_{\text{human}}$($v$)
    \If{$\Delta(v) \geq ~ \tau$ }
        \State $v$ is in the Green List.
    \Else
        \State $v$ is in the Red List.
    \EndIf
\EndFor
\end{algorithmic}
\end{algorithm}

\begin{table}[h]
\centering
  \begin{tabular}{l@{\hspace{0.3em}}c@{\hspace{0.5em}}c@{\hspace{0.5em}}c@{\hspace{0.5em}}c@{\hspace{0.5em}}}
        \toprule
        & Precision & Recall & FPR  & F1 \\
        \midrule
        \makecell{UW}   & 0.89 & 0.48   & 0.05 & 0.62   \\
        \midrule
        \makecell{UW\\generation-based}  & 0.96 & 0.92   & 0.09 & 0.93 \\
        \midrule
        \makecell{SIR\\generation-based} & 0.88 & 0.80   & 0.24 & 0.83 \\
        \bottomrule
    \end{tabular}
  \caption{We report the precision, recall, FPR and F1 for reverse engineering the watermarking using 1 million tokens. This table complements the results reported in Figure \ref{fig:example} and provides additional insight that the difference in F1 score is primarily driven by the difference in recall. This aligns with our intuition that we fail to correctly classify tokens that are less frequent.}
  \label{sec:appendix_other_metrics}
\end{table}

\subsection{Estimating the green list using a different base distribution} \label{sec:appendix_greenlist}

To evaluate the robustness of Algorithm \ref{alg:estimate_green_list}, we present additional results on estimating the green list using the RealNewsLike subset of the c4 dataset \cite{JMLR:v21:20-074} as the base distribution ($D_{\text{human}}$ in Algorithm \ref{alg:estimate_green_list}). The results of these experiments, summarized in Table \ref{table:c4_greenlist_results}, span evaluations on two distinct base models. Our findings demonstrate that the algorithm's performance remains consistent across different choices of base distribution, thus confirming its robustness.

\begin{figure}[ht]
  \centering
  % \includesvg[width=\columnwidth]{figures/proxyF1_other-gamma.svg}
  \includeinkscape[width=\columnwidth]{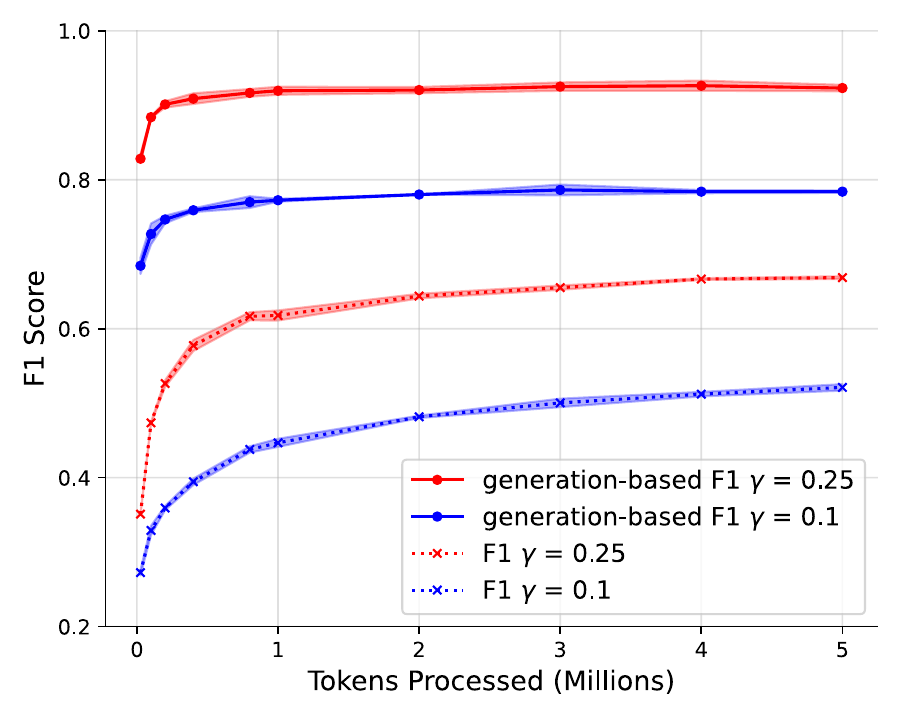}
  \caption{Comparision between the traditional F1 score and generated-based F1 score. We can observe that across all choices of $\gamma$, the traditional F1 metric can understate the security robustness.}
  \label{fig:appendix_gammas}
\end{figure}

\section{Additional Results}
\label{sec:appendix_additional_results}

We present additional results for other choices of $\gamma$ ($0.1$, $0.25$) in  \S\ref{sec:appendix_gamma} and present results on Pythia $1.4$B and Mistral $7$B in  \S\ref{sec:appendix_pythia}. These additional analyses serve to underscore the generalizability of our findings.

\subsection{Results for other choices of $\gamma$} 
\label{sec:appendix_gamma}

We compare the F1 and generation-based F1 score across other choices of $\gamma$ (Figure \ref{fig:appendix_gammas}). We consistently observe a gap between the two metrics. We also note that we can reverse engineer the watermark across all choices of $\gamma$. Additionally, we present the impact of paraphrasing on the watermarking scheme in Table \ref{table:paraphrase_appendix_gammas}. Our results indicate that our attack remains highly effective regardless of the value of $\gamma$.

\begin{table}[ht]
\centering
  \begin{tabular}{l@{\hspace{0.3em}}c@{\hspace{0.5em}}c@{\hspace{0.5em}}c@{\hspace{0.5em}}c@{\hspace{0.5em}}}
        \toprule
        Attack & \makecell{TPR @\\ 1\% FPR} & \makecell{TPR @\\10\% FPR} & P-SP & PPL \\
        \midrule
        No Attack & 100.0 & 100.0 & 1.0 & 15.3  \\
        DIPPER (L20) & 99.2 & 99.8 & 0.95 & 11.8  \\
        DIPPER (L60) & 93.2 & 98.0 & 0.90 & 11.0  \\
        \midrule
        \textbf{Ours (L20)} & 9.0 $\downarrow$ & 19.2 $\downarrow$ & 0.88 & 12.3  \\
        \textbf{Ours (L60)} & 0.2 $\downarrow$ & 0.6 $\downarrow$ & 0.81 & 11.9  \\
        \bottomrule
    \end{tabular}
  \caption{Paraphrasing attacks against EWD algorithm.}
  \label{table:ewd_attack}
\end{table}

\begin{table}[ht]
\centering
  \begin{tabular}{l@{\hspace{0.3em}}c@{\hspace{0.5em}}c@{\hspace{0.5em}}c@{\hspace{0.5em}}c@{\hspace{0.5em}}}
        \toprule
        Attack & \makecell{TPR @\\ 1\% FPR} & \makecell{TPR @\\10\% FPR} & P-SP & PPL \\
        \midrule
        No Attack & 96.5 & 99.7 & 1.0 & 11.6   \\
        DIPPER (L20) & 67.2 & 91.7  & 0.94 & 7.9  \\
        DIPPER (L60) &  43.5 & 82.0 & 0.90 & 7.8   \\
        \midrule
        \textbf{Ours (L20)} & 57.2 $\downarrow$ & 89.5 $\downarrow$ & 0.92 & 8.4  \\
        \textbf{Ours (L60)} & 33.7 $\downarrow$ & 76.0 $\downarrow$ & 0.88 & 8.4  \\
        \bottomrule
    \end{tabular}
  \caption{Paraphrasing attacks against Adaptive Text Watermark.}
  \label{table:awd_attack}
\end{table}

\begin{table*}[ht]
\centering
  \begin{tabular}{lcccccccc}
        \toprule
        & \multicolumn{4}{c}{\unigram with $\gamma=0.1$} & \multicolumn{4}{c}{\unigram with $\gamma=0.25$} \\
        \cmidrule(lr){2-5} \cmidrule(lr){6-9}
        Attack & \makecell{TPR @\\ 1\% FPR} & \makecell{TPR @\\10\% FPR} & P-SP & PPL & \makecell{TPR @\\ 1\% FPR} & \makecell{TPR @\\ 10\% FPR} & P-SP & PPL \\
        \midrule
        No Attack &  97.0 & 99.8 & 1.00 & 13.03 & 99.0 & 100.0 & 1.0 & 15.27 \\
        DIPPER-\textit{L20} & 76.6  & 94.8 & 0.94 & 10.33 & 87.2 & 98.6 & 0.95  & 12.33 \\
        DIPPER-\textit{L60} & 49.2 & 83.6 & 0.90 & 9.85 & 65.2 & 89.8 & 0.90 & 11.05 \\
        \midrule
        \textbf{Ours-\textit{L20}} & 14.0 & 45.6  & 0.90 & 10.68 & 10.6 & 31.6 & 0.87  & 12.2  \\
        \textbf{Ours-\textit{L60}} & 2.8 & 15.8  & 0.85 & 10.20 & 1.6  & 7.6  & 0.81  & 11.4  \\
        \bottomrule
    \end{tabular}
  \caption{Performance of \unigram \cite{zhao2023provable} across different fractions of green list  $\gamma$. We can observe that our attack is highly effective irrespective of the value of $\gamma$.}
  \label{table:paraphrase_appendix_gammas}
\end{table*}

\begin{table*}[ht]
\centering
  \begin{tabular}{lcccccccc}
        \toprule
        & \multicolumn{4}{c}{\unigram} & \multicolumn{4}{c}{\sir} \\
        \cmidrule(lr){2-5} \cmidrule(lr){6-9}
        Attack & \makecell{TPR @\\ 1\% FPR} & \makecell{TPR @\\10\% FPR} & P-SP & PPL & \makecell{TPR @\\ 1\% FPR} & \makecell{TPR @\\ 10\% FPR} & P-SP & PPL \\
        \midrule
        No Attack & 99.4  & 100.0 & 1.0  & 18.7 & 100.0  & 100.0 & 1.0 & 21.1  \\
        DIPPER-\textit{L20} & 96.2 & 99.0 & 94.6 & 13.7 & 87.5 & 98.4 & 0.95 & 15.3  \\
        DIPPER-\textit{L60} & 81.8 & 97.0 & 90.7 & 12.1 & 69.3 & 93.3 & 0.91 & 13.2 \\
        \midrule
        \textbf{Ours -\textit{L20}} & 4.2 & 13.6 & 0.88 & 14.4 & 11.1 & 43.1 & 0.90 & 15.6 \\
        \textbf{Ours -\textit{L60}} & 1.6 & 2.8 & 0.79 & 12.9 & 4.2 & 14.2 & 0.85 & 13.6 \\
        \bottomrule
    \end{tabular}
  \caption{Performance of \unigram and \sir against paraphrasing attacks. Pythia-1.4B is used as the base language model for all the experiments.}
  \label{table:paraphrase_appendix_pythia}
\end{table*}

\begin{table*}[ht]
\centering
  \begin{tabular}{lcccccccc}
        \toprule
        & \multicolumn{4}{c}{\unigram} & \multicolumn{4}{c}{\sir} \\
        \cmidrule(lr){2-5} \cmidrule(lr){6-9}
        Attack & \makecell{TPR @\\ 1\% FPR} & \makecell{TPR @\\10\% FPR} & P-SP & PPL & \makecell{TPR @\\ 1\% FPR} & \makecell{TPR @\\ 10\% FPR} & P-SP & PPL \\
        \midrule
        No Attack & 98.4  & 99.8 & 1.0 & 15.8 & 83.3 & 93.5 & 1.0 & 11.9   \\
        DIPPER-\textit{L20} & 87.4 & 98.0 & 0.95 & 11.8 & 48.0 & 72.2 & 0.95 & 9.6  \\
        DIPPER-\textit{L60} & 64.6 & 91.4 & 0.91 & 10.7 & 25.5 & 52.8 & 0.92 & 9.2 \\
        \midrule
        \textbf{Ours -\textit{L20}} & 5.4  & 30.9  & 0.89 & 12.5 & 1.3 & 6.6 & 0.9 & 10.4 \\
        \textbf{Ours -\textit{L60}} & 0.4  &  4.0  & 0.82 & 11.7 & 0.2 & 1.1 & 0.82 & 10.4 \\
        \bottomrule
    \end{tabular}
  \caption{Performance of \unigram and \sir against paraphrasing attacks. Mistral 7B is used as the base language model for all the experiments.}
  \label{table:paraphrase_appendix_mistral}
\end{table*}

\subsection{Results on Additional Models}
\label{sec:appendix_pythia}

We present the performance of watermarking schemes against paraphrasing attacks, using Pythia-1.4b \cite{biderman2023pythia} and Mistral 7B \cite{jiang2023mistral7b} as the base language models in Table \ref{table:paraphrase_appendix_pythia} and 
Table~\ref{table:paraphrase_appendix_mistral}. 
For all experiments, we set the watermark hyperparameters to $\gamma=0.5$ and $\delta=2.0$. Our results demonstrate that the proposed paraphrasing attack significantly degrades the performance of both watermarking schemes evaluated. These results indicate that our findings are applicable across different model classes and sizes.

\subsection{Efficacy of our paraphrasing attack on prompts from diverse datasets}
\label{sec:appendix_additional_datasets}

To demonstrate the effectiveness of our proposed paraphrasing attack on generations from diverse domains, we evaluate our attack on prompts from arXiv papers \cite{cohan2018discourseawareattentionmodelabstractive} and Booksum \cite{kryściński2022booksumcollectiondatasetslongform}. We follow a similar setup as explained in  \S\ref{exp:dataset}. These results also serve as evidence that the green list estimated once using a particular dataset (OpenWebText \cite{Gokaslan2019OpenWeb} in this case) can be used to launch paraphrasing attacks on a variety of downstream datasets. The results are summarized in Table \ref{table:appendix_arxiv} and Table \ref{table:appendix_booksum} for arXiv papers and Booksum dataset respectively.

\section{Paraphrasing attacks against EWD}
\label{sec:entropy_based_text_watermarking}

Entropy-based Text Watermarking Detection (EWD) \cite{lu2024entropybasedtextwatermarkingdetection} introduces a novel approach to watermark detection by incorporating token entropy. This method assigns higher importance to high-entropy tokens during the detection process, thereby enhancing detection performance in low-entropy contexts. We conducted an empirical investigation into the robustness of EWD against paraphrasing attacks. 
The results 
of our
analysis 
are presented in 
Table \ref{table:ewd_attack}, 
providing insights 
into the method's resilience to 
paraphrasing attacks. 
From the results 
we can observe that incorporating 
the estimated green list 
can significantly improve 
the 
effectiveness 
of the paraphrasing attack.

\begin{table*}[h]
\centering
  \begin{tabular}{lcccccccc}
        \toprule
        & \multicolumn{4}{c}{\unigram} & \multicolumn{4}{c}{\sir} \\
        \cmidrule(lr){2-5} \cmidrule(lr){6-9}
        Attack & \makecell{TPR @\\ 1\% FPR} & \makecell{TPR @\\10\% FPR} & P-SP & PPL & \makecell{TPR @\\ 1\% FPR} & \makecell{TPR @\\ 10\% FPR} & P-SP & PPL \\
        \midrule
        No Attack & 100 & 100 & 1 & 27.9 & 98.6 & 100.0 & 1.0  & 29.6  \\
        DIPPER (L20) & 93.4 & 99.2 & 0.93 & 19.87 & 65.3 & 87.7 & 0.94 & 22.0 \\
        DIPPER (L60) & 57.2 & 88.8 & 0.88 & 16.3 & 23.1 & 59.3 & 0.88 & 16.80 \\
        \midrule
        \textbf{Ours (L20)} & 2.4 $\downarrow$ & 11.2 $\downarrow$ & 0.85 & 19.1 & 2.6 $\downarrow$ & 11.7 $\downarrow$ & 0.86 & 17.87 \\
        \bottomrule
    \end{tabular}
    \caption{Result demonstraing the efficacy of paraphrasing attacks on prompts from the arXiv papers dataset.}
  \label{table:appendix_arxiv}
\end{table*}

\begin{table*}[h]
\centering
  \begin{tabular}{lcccccccc}
        \toprule
        & \multicolumn{4}{c}{\unigram} & \multicolumn{4}{c}{\sir} \\
        \cmidrule(lr){2-5} \cmidrule(lr){6-9}
        Attack & \makecell{TPR @\\ 1\% FPR} & \makecell{TPR @\\10\% FPR} & P-SP & PPL & \makecell{TPR @\\ 1\% FPR} & \makecell{TPR @\\ 10\% FPR} & P-SP & PPL \\
        \midrule
        No Attack & 99.8 & 100.0 & 1.0 & 28.1 & 99.3 & 99.7 & 1.0 & 25.0  \\
        DIPPER (L20) & 99.6 & 100.0 & 0.94 & 24.4 & 91.7 & 74 & 0.94  & 23.8  \\
        DIPPER (L60) & 92.0 & 98.4 & 0.87 & 22.5 & 39.1 & 68.2 & 0.88 & 20.6 \\
        \midrule
        \textbf{Ours (L20)} & 9.8 $\downarrow$ & 26.4  $\downarrow$ & 0.83 & 24.61 & 16.8 $\downarrow$ & 32.4 $\downarrow$ & 0.87 & 22.9 \\
        \bottomrule
    \end{tabular}
    \caption{Result demonstraing the efficacy of paraphrasing attacks on prompts from the Booksum dataset.}
  \label{table:appendix_booksum}
\end{table*}

\end{document}